\documentclass[aip,graphicx]{revtex4-1}
\usepackage{amsmath,amsfonts,amssymb}
\usepackage{chemformula} 
\usepackage{graphicx}
\usepackage{listings}
\usepackage{booktabs}
\usepackage{multirow}
\usepackage{siunitx}
\usepackage[flushleft]{threeparttable}
\draft 
\AtBeginDocument{%
	\heavyrulewidth=.08em
	\lightrulewidth=.05em
	\cmidrulewidth=.03em
	\belowrulesep=.65ex
	\belowbottomsep=0pt
	\aboverulesep=.4ex
	\abovetopsep=0pt
	\cmidrulesep=\doublerulesep
	\cmidrulekern=.5em
	\defaultaddspace=.5em
}
\begin{document}


\title{A Rotational/Roto-translational Constraint Method for Condensed Matter} 



\author{Jitai Yang}  
\affiliation{Institute of Theoretical Chemistry, College of Chemistry, Jilin University, 2519 Jiefang Road,
	Changchun 130023, P.R.China}
\author{Ke Li}  
\affiliation{Institute of Theoretical Chemistry, College of Chemistry, Jilin University, 2519 Jiefang Road,
	Changchun 130023, P.R.China}
\author{Jia Liu}  
\affiliation{Institute of Theoretical Chemistry, College of Chemistry, Jilin University, 2519 Jiefang Road,
	Changchun 130023, P.R.China}
\author{Jia Nie}  
\affiliation{Institute of Theoretical Chemistry, College of Chemistry, Jilin University, 2519 Jiefang Road,
	Changchun 130023, P.R.China}

\author{Hui Li}
\email{prof{\textunderscore}huili@jlu.edu.cn}
\affiliation{Institute of Theoretical Chemistry, College of Chemistry, Jilin University, 2519 Jiefang Road,
	Changchun 130023, P.R.China}



\date{\today}

\begin{abstract}
Molecular rotations influence numerous condensed matter phenomena but are often difficult to isolate in MD simulations.  This work presents a rotational/roto-translational constraint algorithm designed for condensed matter simulations.  The method is based on the velocity Verlet scheme, ensuring a direct constraint on velocity and simplifying implementation within material simulation software packages.
We implemented the algorithm in a customized version of CP2K package and validated its effectiveness through MD simulations of molecule and crystal. The results demonstrate successful selective constraint of rotational and roto-translational motions, enabling stable long-term simulations.  This capability opens avenues for studying rotation-related phenomena (e.g., paddle-wheel mechanism in solid-state electrolytes) and constrained sampling.
\end{abstract}

\maketitle 

\section{Introduction}
\label{intro}
Molecular rotations within condensed matter play a significant role in various phenomena. For example, it drives phase transitions in amorphous ice.\cite{rotPhase}
Similarly, the interplay between volumetric lattice strain and the rotational degrees of freedom (DOF) of \ch{CH3NH3} molecules is believed to influence polaron creation and stability in perovskite solar cell materials.\cite{rotSolar}
Furthermore, anion rotation in solid electrolytes is associated with fast ion conduction, and the coupling between this rotation and cation diffusion, known as the paddle-wheel mechanism, has attracted considerable research interest.\cite{Zhang2022, sse1, sse2, yrHu, matter2020, fixMD2}

Experimental techniques offer statistically averaged information of  these rotations. However, combining them with molecular dynamics (MD) simulations allows for more detailed, microscopic insights.
While MD simulations can estimate the effect of rotation on material properties, they often achieve this by fixing the translational, rotational and vibrational motions of the molecules all together.\cite{fixMD1, fixMD2, fixMD3} This approach is limiting, as a more comprehensive understanding requires isolating the effects of rotation alone.

Several research groups have explored constraint algorithms to selectively fix the rotational motion on molecules. The Nola group developed a roto-translational constraint method based on the leapfrog scheme.\cite{leapFrog}
The leapfrog scheme, however, presents limitations in handling velocities compared to the widely used velocity Verlet scheme.\cite{Allen2017}
The Krimm group proposed a general constraint algorithm named WIGGLE, which utilizes constrained accelerations.\cite{wiggle} While WIGGLE boasts high accuracy, it's rarely implemented in popular MD software packages.
 Additionally, the Liu group presented a constraint method based on a ``middle'' scheme, which also deviates from the velocity Verlet scheme.

To our knowledge, there is currently no published implementation of a rotational constraint algorithm based on the widely used velocity Verlet scheme within popular material simulation software. This work presents a rotational/roto-translational constraint method specifically designed for condensed matter simulations. It leverages the advantages of the velocity Verlet integrator, ensuring a direct constraint on velocity updates and simplifying implementation within material simulation software packages. We have successfully implemented this method within a customized version of the CP2K software package.\cite{cp2kShort} The effectiveness of this method will be demonstrated through detailed MD simulations, the results of which will be presented in subsequent sections.
\section{Theory}
\label{theory}
The theoretical derivation of the rotational constraints using the velocity Verlet integrator is presented in this section. Our derivation follows the RATTLE algorithm,\cite{rattle} which is widely applied in MD software.
Constraints for linear molecules are special cases and are not discussed in this work.
Since the roto-translational constraint differ only in the initial conditions from the rotational constraint, the derivation of roto-translational constraint is presented in the supporting information.

We describe the constrained atoms using the center-of-mass (COM) coordinates system for convenience:
\begin{equation}
	\begin{aligned}
		&\mathbf{r}_a = \mathbf{R}_a -\frac{\sum_{a=1}^{n_a} m_a\mathbf{R}_a }{\sum_{a=1}^{n_a}m_a},\\
		&\sum_{a=1}^{n_a} m_a \mathbf{r}_a=\mathbf{0}.
	\end{aligned} \label{eq:comCondition}
\end{equation}
$n_a$ is the total number of constrained atoms, $m_a$ is the mass of $a$th atom. $\mathbf{R}_a$ is the position vector of $a$th atom in Cartesian coordinates system, while $\mathbf{r}_a$ is the position vector of $a$th atom in the COM coordinates system.

We use the same rotational constraint conditions as Nola  groups.\cite{rotc2000}
\begin{equation}
	\boldsymbol\chi^{(r)} = \sum^{n_a}_{a=1} \mathbf r^0_a \times m_a \mathbf r_a(t) = \mathbf 0 \label{eq:consR}.
\end{equation}
The superscript $r$ indicates the quantity corresponds to a positional constraint, and $\mathbf{r}_a^0$ is the rotational reference position of $a$th atom.
Another option is to use condition equations without mass. 
We prefer the condition equations with mass because it is close to the rotational Eckart conditions which divide the nuclear motion into translations, rotations, and vibrations approximately.\cite{Eckart}

The constraints are applied to the whole molecule, some bond wiggling is allowed. In particular, if some atom's reference position is at the COM ($\mathbf{r}_a^0 = \mathbf{0}$), the atom will be conunted in the COM and has no contribution to the summation in Eq.~\ref{eq:consR}.

Differentiating the equations at time $t$ yields the velocity constraint condition
\begin{equation}
	\boldsymbol\chi^{(v)} = \sum^{n_a}_{a=1} \mathbf r^0_a \times m_a \mathbf v_a(t) = \mathbf 0.
	\label{eq:consV}
\end{equation}
The superscript $v$ indicates the quantity corresponds to a velocity constraint, $\mathbf{v}_a(t)$ is the velocity of $a$th atom at time $t$. The rotational reference positions are constants. If we replace $\mathbf{r}_a^0$ with $\mathbf{r}_a(t)$, $\boldsymbol{\chi}^{(v)}$ will be  the total angular moment of the constrained atoms.

Unlike the previous work,\cite{rotc2000} our method employs the widely used velocity Verlet scheme, aligning with the dominant approach in material simulations. Most popular material MD software packages, like LAMMPS,\cite{lammps} VASP,\cite{vasp0, vasp1} and CP2K,\cite{cp2kShort} rely on the velocity Verlet scheme. This compatibility will simplify the implementation into existing software functionalities. Furthermore, the velocity Verlet scheme inherently allows for direct constraint on velocity updates.

The velocity Verlet integrator has position part and velocity part
\begin{equation}
	\begin{aligned}
		& \mathbf{r}_a(t+\delta t)=\mathbf{r}_a(t)+\delta t \,\mathbf{v}_a(t)+\frac{1}{2 m_a} \delta t^2 \mathbf{f}_a(t), \\
		& \mathbf{v}_a(t+\delta t)=\mathbf{v}_a(t)+\frac{1}{2 m_a} \delta t\left(\mathbf{f}_a(t)+\mathbf{f}_a(t+\delta t)\right).
	\end{aligned}
\end{equation}
Consider the position part with the constraint force first
\begin{equation}
	\mathbf{r}_a(t+\delta t)=\mathbf{r}_a^{\prime}(t+\delta t)+\frac{1}{2m_a}\delta t^2  \mathbf{g}_a^{(\mathrm{r})}(t).
	\label{eq:intr}
\end{equation}
$\mathbf{g}_a^{(\mathrm{r})}$ is the positional constraint force on $a$th atom, the prime symbol in $\mathbf{r}_a^\prime$ indicates the quantities without constraints.
Multiply both sides of the equation by $m_a$, cross product both sides of the equation by $\mathbf{r}_a^0$ from left, then sum over atoms, we get
\begin{equation}
	\sum^{n_a}_{a=1}  \mathbf r_a^0 \times m_a\mathbf{r}_a(t+\delta t)=
	\sum^{n_a}_{a=1} \mathbf r_a^0 \times 
	[ m_a\mathbf{r}_a^{\prime}(t+\delta t)+\frac{1}{2}\delta t^2 \, \mathbf{g}_a^{(\mathrm{r})}(t)].
	\label{eq:consrMid0}
\end{equation}
The left side of the equation is the same as our rotational  constraint conditions Eq.~(\ref{eq:consR}), which should be zero at anytime. So
\begin{equation}
	\sum^{n_a}_{a=1} \mathbf r_a^0 \times 
	[ m_a\mathbf{r}_a^{\prime}(t+\delta t)+\frac{1}{2}\delta t^2 \, \mathbf{g}_a^{(\mathrm{r})}(t)] = \mathbf 0.
	\label{eq:consrMid}
\end{equation}
The positional constraint force can be written as 
\begin{equation}
	\mathbf{g}_a^{(r)}=-\sum_{i=1}^{n_r} \lambda_i^{(r)} \nabla_{\!\mathbf{r}_a}\, \chi_i^{(r)}. \label{eq:gr0}
\end{equation}
$n_r$ is the number of DOF are constrainted, $n_r =3$ in rotational constraint for nonlinear molecule.
$\lambda_i^{(r)}$ is the $i$th Lagrange multipliers of position constraint. The minus symbol on the right side is just for definition because we can write it into the $\lambda_i^{(r)}$. 

Substitute Eq.~(\ref{eq:consR}) into Eq.~(\ref{eq:gr0}). Take $(x_1, y_1, z_1, x_2, y_2, z_2, \dots, x_a,y_a,z_a)$ as atoms position in COM coordinate system, $(X_1, Y_1, Z_1, X_2, Y_2, Z_2, \dots, X_a,Y_a,Z_a)$ as atoms position of rotational reference in COM coordinate system. We get
\begin{equation}
	\begin{aligned}
		&g_{a,x}^{(r)} = -\sum_{i=1}^{n_r} \lambda_i^{(r)} \frac{\partial \chi_i^{(r)}}{\partial x_a}
		=m_a\left(- Z_a\lambda_2^{(r)}+ Y_a\lambda_3^{(r)}\right), \\
		&g_{a,y}^{(r)} = -\sum_{i=1}^{n_r} \lambda_i^{(r)} \frac{\partial \chi_i^{(r)}}{\partial y_a}
		=m_a\left(\ \ Z_a\lambda_1^{(r)} -X_a\lambda_3^{(r)} \right), \\
		& g_{a,z}^{(r)} = -\sum_{i=1}^{n_r} \lambda_i^{(r)} \frac{\partial \chi_i^{(r)}}{\partial z_a}
		=m_a\left(-Y_a\lambda_1^{(r)} +X_a\lambda_2^{(r)} \right).
	\end{aligned}
	\label{eq:gr}
\end{equation}
Define
\begin{equation}
	\mathbf{c}^{(r)}= \sum_{a=1}^{n_a} \mathbf{r}_a^0 \times m_a\mathbf{r}_a^\prime(t+\delta t). \label{eq:cr}
\end{equation}
Substitute Eq.~(\ref{eq:gr}) and Eq.~(\ref{eq:cr}) into Eq.~(\ref{eq:consrMid}), we will get equations
\begin{equation}
	\begin{cases}
		\mathbf{c}_x^{(r)} +\sum\limits_{a=1}^{n_a} \frac{1}{2} m_a\delta t^2
		\left( -\left(Y_a^2 +Z_a^2\right) \lambda_1^{(r)}
		+ X_aY_a \lambda_2^{(r)}
		+ X_aZ_a \lambda_3^{(r)}\right) =0 \\
		
		\mathbf{c}_y^{(r)} +\sum\limits_{a=1}^{n_a} \frac{1}{2} m_a\delta t^2
		\left(\ \ \ X_aY_a \lambda_1^{(r)}
		-\left(X_a^2 +Z_a^2\right) \lambda_2^{(r)}
		+ Y_aZ_a \lambda_3^{(r)}\right)  = 0 \\
		
		\mathbf{c}_z^{(r)} +\sum\limits_{a=1}^{n_a} \frac{1}{2} m_a\delta t^2
		\left(\ \ \, X_a Z_a \lambda_1^{(r)}
		+Y_aZ_a \lambda_2^{(r)}
		-\left(X_a^2 +Y_a^2\right)  \lambda_3^{(r)}\right) = 0.
	\end{cases}
\end{equation}
In matrix form is
\begin{equation} 
	\begin{aligned}
		&\mathbf{c}^{(r)} + \boldsymbol{\Theta}^{(r)}\boldsymbol\lambda^{(r)}= \mathbf{0},\\
		&\boldsymbol{\lambda}^{(r)}= 
		-\left(\boldsymbol{\Theta}^{(r)} \right)^{-1} \mathbf{c}^{(r)}.
	\end{aligned} \label{eq:consrSolve1}
\end{equation}
Then we can solve all Lagrange multipliers $ \boldsymbol\lambda^{(r)} $. From Eq.~(\ref{eq:gr}) and Eq.~(\ref{eq:intr}) we can get constraint force and constrained positions at the next time step $\mathbf{r}_a(t+\delta t )$.

Within the velocity Verlet scheme, it is straightforward to incorporate constraints on velocity updates. The velocity part of velocity Verlet integrator with constraint force is
\begin{equation}
	\mathbf{v}_a(t+\delta t)=\mathbf{v}_a^{\prime}(t+\delta t)+\frac{1}{2m_a}\delta t \, \mathbf{g}_a^{(\mathrm{v})}(t+\delta t).\label{eq:consvMid}
\end{equation}
$ \mathbf{g}_a^{(\mathrm{v})} $ is the velocity constraint force on $a$th atom. As we have done in the position part, we multiply both sides of the equation by $m_a$, cross product both sides of the equation by $\mathbf{r}_a^0$ from left, then sum over atoms
\begin{equation}
	\sum_{a=1}^{n_a} \mathbf{r}_a^0 \times m_a \mathbf{v}_a(t+\delta t)
	=\sum_{a=1}^{n_a} \mathbf{r}_a^0 \times\left[m_a \mathbf{v}_a^{\prime}(t+\delta t)+\frac{1}{2}\delta t\, \mathbf{g}_a^{(\mathrm{v})}(t + \delta t)\right].
	\label{eq:consvMid0}
\end{equation}
From Eq.~(\ref{eq:consV})
\begin{equation}
	\sum_{a=1}^{n_a} \mathbf{r}_a^0 \times[ m_a
	\mathbf{v}_a^{\prime}(t+\delta t)+\frac{1}{2}\delta t\, \mathbf{g}_a^{(\mathrm{v})}(t+\delta t)]
	= \mathbf  0.
	\label{consVEnd}
\end{equation}
Eq.~(\ref{eq:consV}) is a semi-holonomic constraint condition.  The corresponding constraint force is\cite{Goldstein2013}
\begin{equation}
	\mathbf{g}_a^{(v)} = - \sum ^{n_v} _{i=1}\lambda _i^{(v)} \boldsymbol{\nabla}_{\!\mathbf v_a} 
	\,\chi_i^{(v)} \label{eq:consFV}
\end{equation}
Note that the gradient is with respect to the velocity $\mathbf{v}_a$ rather than positions. The rotational reference positions are constants, so the components of $ \mathbf{g}_a^{(\mathrm{v})} $ has the same formalism as $ \mathbf{g}_a^{(\mathrm{r})} $ in Eq.~(\ref{eq:gr})
\begin{equation}
	\begin{aligned}
		&g_{a,x}^{(v)} = -\sum_{i=1}^{n_v} \lambda_i^{(v)} \frac{\partial \chi_i^{(v)}}{\partial v_{a,x}}
		=m_a\left(- Z_a\lambda_2^{(v)}+ Y_a\lambda_3^{(v)}\right), \\
		&g_{a,y}^{(v)} = -\sum_{i=1}^{n_v} \lambda_i^{(v)} \frac{\partial \chi_i^{(v)}}{\partial v_{a,y}}
		=m_a\left(\;\;Z_a\lambda_1^{(v)} -X_a\lambda_3^{(v)} \right), \\
		& g_{a,z}^{(v)} = -\sum_{i=1}^{n_v} \lambda_i^{(v)} \frac{\partial \chi_i^{(v)}}{\partial v_{a,z}}
		=m_a\left(-Y_a\lambda_1^{(v)} +X_a\lambda_2^{(v)} \right).
	\end{aligned}\label{eq:gv}
\end{equation}
Define
\begin{equation}
	\mathbf{c}^{(v)}= \sum_{a=1}^{n_a} \mathbf{r}_a^0 \times m_a\mathbf{v}_a^\prime(t+\delta t).\label{eq:cv}
\end{equation}
Substitute Eq.~(\ref{eq:gv}) and Eq.~(\ref{eq:cv}) into Eq.~(\ref{consVEnd}) we will get 
\begin{equation}
	\begin{cases}
		\mathbf{c}_x^{(v)} +\sum\limits_{a=1}^{n_a} \frac{1}{2} m_a\delta t
		\left( -\left(Y_a^2 +Z_a^2\right) \lambda_1^{(v)}
		+ X_aY_a \lambda_2^{(v)}
		+ X_aZ_a \lambda_3^{(v)}\right) =0 \\
		
		\mathbf{c}_y^{(v)} +\sum\limits_{a=1}^{n_a} \frac{1}{2} m_a\delta t
		\left( \quad X_aY_a \lambda_1^{(v)}
		-\left(X_a^2 +Z_a^2\right) \lambda_2^{(v)}
		+ Y_aZ_a \lambda_3^{(v)}\right)  = 0 \\
		
		\mathbf{c}_z^{(v)} +\sum\limits_{a=1}^{n_a} \frac{1}{2} m_a\delta t
		\left( \quad\! X_a Z_a \lambda_1^{(v)}
		+Y_aZ_a \lambda_2^{(v)}
		-\left(X_a^2 +Y_a^2\right)  \lambda_3^{(v)}\right) = 0.
	\end{cases}
\end{equation}
In matrix form is
\begin{equation}
	\begin{aligned}
		&\mathbf{c}^{(v)} + \boldsymbol{\Theta}^{(v)}\boldsymbol\lambda^{(v)}= \mathbf{0},\\
		&\boldsymbol{\lambda}^{(v)}= 
		-\left(\boldsymbol{\Theta}^{(v)} \right)^{-1} \mathbf{c}^{(v)}.
		\label{eq:consrVend}
	\end{aligned}
\end{equation}
Then $\mathbf{\lambda}_a^{(v)}, \,\mathbf{g}_a^{(v)}$ and the next step velocity $\mathbf{v}_a(t+\delta t)$  are solved. 

Using inverse matrix method to solve the Lagrange multipliers is usually considered to be slow. However, no matter how many atoms are in one rotational constraint, the constrained DOF are only 3 (6 for roto-translational constraint); we only need to ask for the inverse of two $3\times3$ ($6\times 6$ for roto-translational constraint) matrices. And the more efficient elimination method can be used too.

At the next step, we compute all the Lagrange multipliers, constrained positions, and constrained velocities again, then loop until the end of MD. 
\section{Molecule dynamics results }
\label{res}
\subsection{Computational details}
The running details of the MD trajectory results are as follows: 

\textbf{Three ammonia molecules}: The system consists of three ammonia molecules, as in Fig.~\ref{fig:snapshots}.
The HFX (Hartree-Fock exchange) basis set\cite{hfx1, hfx2} and the GTH pseudopotentials\cite{gth1} are used in the AIMD simulation. The simulation box is a cube of 10~\AA\ side length, periodic boundary condition (PBC) is used. In force field molecular dynamics (FFMD), Amber format force field\cite{amber22} and 25~\AA\ side length box with PBC is used for the modeling. 
The initial temperature is 300 K. 
The time step is 1 fs, and the length of each trajectory is 2 ps.
The NVE and the NVT ensemble are used. The NOS\'{E} thermostator\cite{nose1, nose2} is used in the NVT ensemble, and the temperature is set as 300~K. 
Only AIMD results are shown below. FFMD results are in the Supporting Information.
Then follow the equillibrium trajectory by doing unconstrained, rotational constrained, and roto-translational constrained FFMD for 1 ns each.

\begin{figure}[hbtp]
	\centering
	\includegraphics{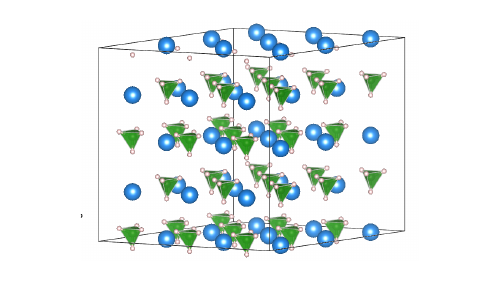} \\
	\caption{The structure of crystal \ch{LiBH4}.
		\label{fig:mol_crys}}
\end{figure}

\textbf{\ch{LiBH4} crystal}: The system consists of the \ch{LiBH4} crystal as showed in Fig.~\ref{fig:mol_crys}. A $3\times3\times2$ suppercell is made for simulation, the unitcell structure is from the Yvon group's work.\cite{struLiBH4}
 The DZVP-MOLOPT-SR-GTH\cite{cp2kGaussBasis} and PBE GTH pseudopotentials\cite{gth1996, gth1998, gth1} are used for AIMD. The time step is 0.5 fs, the unconstrained, rotational constrained and roto-translational constrained trajectories are of length 10 ps, the last 5 ps trajectories are for statistic of conserved quantity in Fig. \ref{fig:libh4Comp}. NVT ensemble and NOS\'{E} thermostator is used, the temperature is set as 800 K.

All MD results ran with customized version codes of CP2K v8.2.
\subsection{MD results}
\subsubsection{Three ammonia molecules}
\begin{figure}[hbtp]
	\centering
	\includegraphics{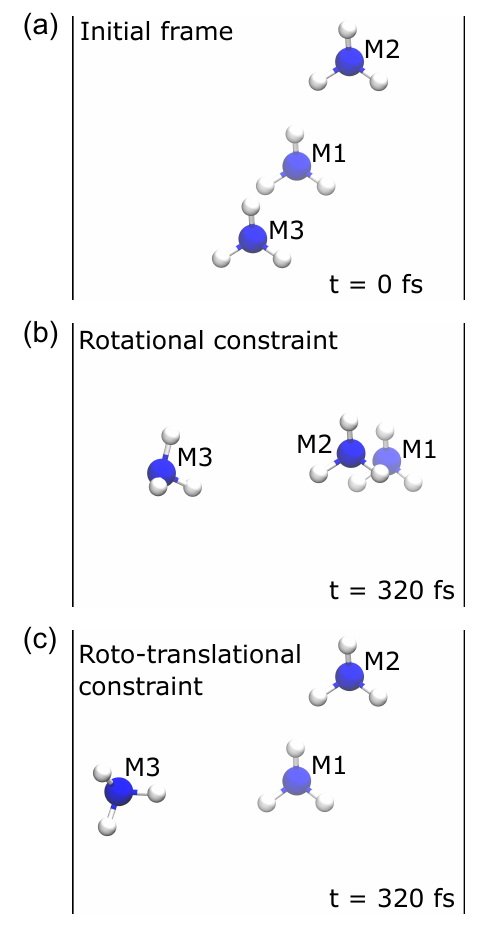} \\
	\caption{Snapshots of AIMD trajectory of three ammonia molecules system: (a) The init frame of AIMD; (b) The snapshot of trajectory using rotational constraint at 320~fs. Molecules M1 and M2 are constrained; (c) The snapshot of trajectory using roto-translational constraint at 320~fs. Molecules M1 and M2 are constrained. The animations of trajectories are in Supporting Information.
		\label{fig:snapshots}}
\end{figure}

\textbf{Rotational constraint results}:
The snapshot of NVE ensemble AIMD trajectory using rotational constraints are shown in Fig.~\ref{fig:snapshots}(a) (t = 0 fs) and Fig.~\ref{fig:snapshots}(b) (t = 320 fs).
One can check animations in the Supporting Information. We define rotational constraint error as the norm of deviation from Eq.~(\ref{eq:consR}) 
\begin{equation}
	\sigma =  \left\| \sum^{n_a}_{a=1} \mathbf r^0_a \times m_a \mathbf r_a(t)\right\| .\label{eq:consER}
\end{equation}
In AIMD the max error is $\sigma_{\max} = 9.43 \times 10^{-7}$~\AA, and in FFMD the max error is $\sigma_{\max} = 1.52 \times 10^{-6}$~\AA, which are close to numerical errors. The result shows the molecule M3 moves normally, and the other two constrained molecules M1 and M2 do not rotate while translating and vibrating, which means our program meets the expectation.

\begin{figure}[hbtp]
	\centering
	\includegraphics[width=.5\textwidth]{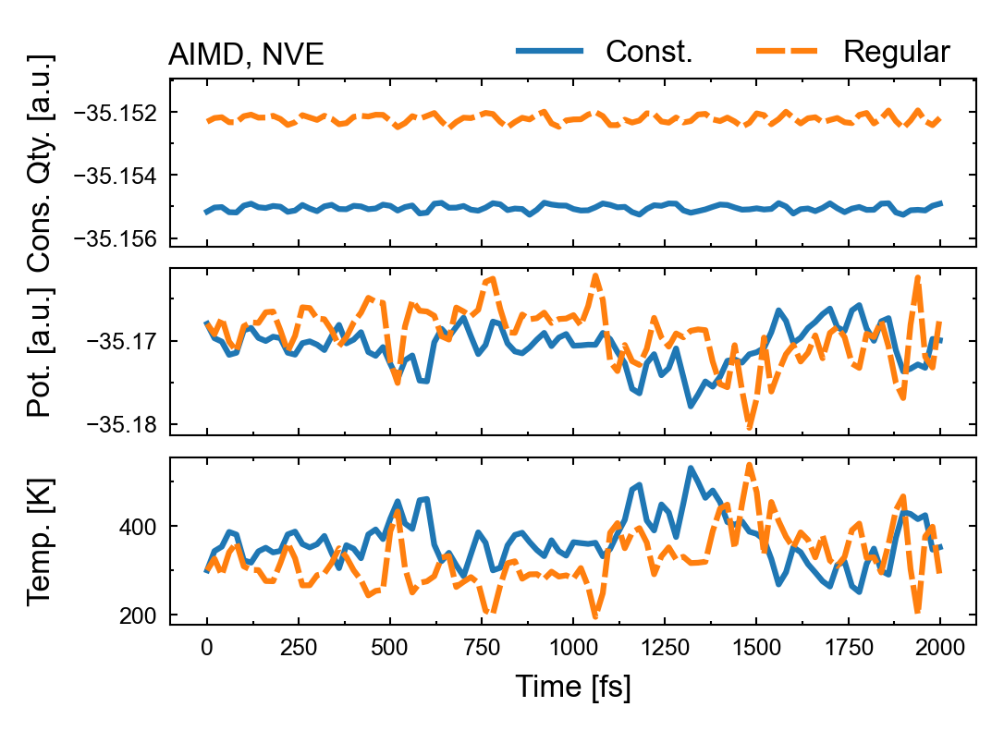} \\
	\caption{Conserved quantity (Cons.~Qty.), potential (Pot.), and temperature (Temp.) fluctuation of the AIMD trajectory of NVE ensemble with/without rotational constraints (Const./Regular). \label{fig:rc_qm_nve}}
\end{figure}

\begin{figure}[hbtp]
	\centering
	\includegraphics[width=.5\textwidth]{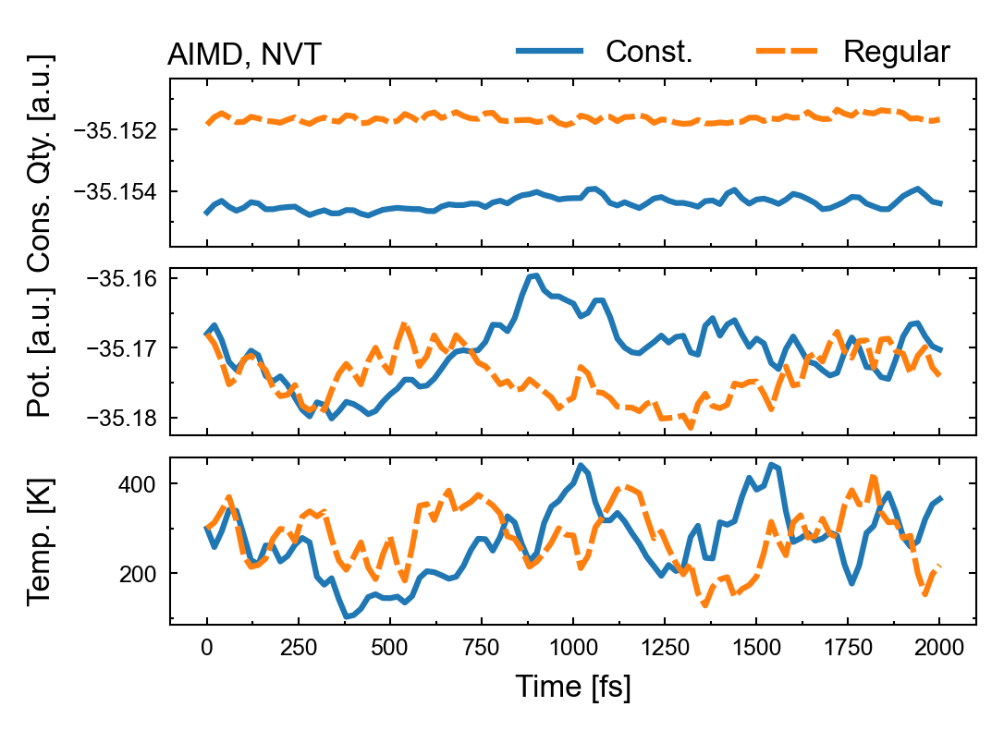} \\
	\caption{Conserved quantity (Cons.~Qty.), potential (Pot.), and temperature (Temp.) fluctuation of the AIMD trajectory of NVT ensemble with/without rotational constraints (Const./Regular). \label{fig:rc_qm_nvt}}
\end{figure}

One important question is the numerical stability of the program and insuring the errors will not accumulate during the running of MD. We statistic the fluctuations of conserved quantity, potential and temperature in trajectories, then compare the data with MD without constraints. The results of AIMD are shown in Fig.~\ref{fig:rc_qm_nve}--\ref{fig:rc_qm_nvt}. 
More results of comparison of FFMD with/without constraints are in Supporting Information. 
The constant quantity (Cons. Qty.) indicates the number stability of the MD.
It corresponds to the total energy in NVE ensemble, and in the NVT ensemble, it corresponds to the total energy of the system containing DOF of the heat bath. The difference of conserved quantity with/without constraints comes from the DOF change by constraints. The results show that no significant difference in the fluctuation size of conserved quantity, potential and temperature.
We also computed the drift and fluctuation of conserved quantity by least-squares fitting. We performed a least-squares fit of data to a straight line respectively. The drift is defined as the difference of the fit at the first and last point. The fluctuation is defined as the root mean square deviation (RMSD) around the least-squares fit. 

The amplitude of fluctuation in potential and conserved quantity is large when compares with the average values. This is because we chose a  small system for test, we will discuss a larger example of \ch{LiBH4} crystal system in the section\,\ref{sec:libh4}.

All drift/fluctuation difference of conserved quantity from the no constraint MD is showed in Table~\ref{tbl:consq}. 
Most drift and fluctuation difference values are little than $1\times10^{-4}$~a.u. The drift difference in AIMD using NVT ensemble showed a little bigger result with $2.57\times10^{-4}$~a.u., which may caused by the collision of two ammonia molecules in the trajectory.

\begin{table}[]
	\begin{threeparttable}
		\begin{tabular}{lllS[table-number-alignment=right]S[table-number-alignment=right]S[table-number-alignment=right]S[table-number-alignment=right]S[table-number-alignment=right]}
			\toprule
			&&&
	\multicolumn{2}{c}{Molecule(\ch{NH3})}&
	\multicolumn{2}{c}{Crystal(\ch{LiBH4})}\\
			\midrule
			MD type                      & Ensemble             &       & \text{$\Delta_\text{RC}\,^\text{a}$}   & \text{$\Delta_\text{RTC}\,^\text{b}$}&\text{$\Delta_\text{RC}$}   & \text{$\Delta_\text{RTC}$} \\ \midrule
			\multirow{4}{*}{Ab initio}   & \multirow{2}{*}{NVE} & Drift & 0.01 & 0.10 &&\\
			&                      & Fluc. & 0.36 & 0.26 &&\\
			& \multirow{2}{*}{NVT} & Drift & 2.57 & 0.11 &2.24&1.87\\
			&                      & Fluc. & 0.58 & 0.50 &0.67&0.43\\
			\multirow{4}{*}{Force field} & \multirow{2}{*}{NVE} & Drift & 0.03 & 0.02 &&\\
			&                      & Fluc. & 0.07 & 0.06 &&\\
			& \multirow{2}{*}{NVT} & Drift & 0.13 & 0.07 &&\\
			&                      & Fluc. & 0.12 & 0.10 &&\\
			\bottomrule
		\end{tabular}
		
		\begin{tablenotes}
			\small
			\item $\,^\text{a}$Rotational constraint
			\item $\,^\text{b}$Roto-translational constraint
			\item *The unit of data is $1\times 10^{-4}$~a.u. 
		\end{tablenotes}
	\end{threeparttable}\caption{Drift and fluctuation (Fluc.) difference of conserved quantity (unit: $1\times 10^{-4}$~a.u.) between constrained MD and regular MD, which indicates the numerical stability.}\label{tbl:consq}
\end{table}

\textbf{Roto-translational constraint results}:
The snapshot of NVE ensemble AIMD trajectory using roto-translational constraints are shown in Fig.~\ref{fig:snapshots}(a) (t = 0 fs) and Fig.~\ref{fig:snapshots}(c) (t = 320 fs).
One can check animations in the Supporting Information for more direct view. We define the rotational constraint error same as in Eq.~(\ref{eq:consER}). In AIMD the max rotational error is $\sigma_{\max} = 3.54 \times 10^{-7}$~\AA, and in FFMD the max rotational error is $\sigma_{\max} = 4.83 \times 10^{-6}$~\AA. 
The translational error is defined as $\tau = \left\| \mathbf{C}^0 - \mathbf{C}(t)\right\|$, which is the vector norm of change of center-of-mass $\mathbf{C}$. In AIMD the max translational error is $\tau_{\max} = 1.00 \times 10^{-6}$~\AA, and in FFMD the max translational error is $\tau_{\max} = 5.04 \times 10^{-7}$~\AA, , which are close to numerical error. The result shows the molecule M3 moves normally, and the other two constrained molecules M1 and M2 do not rotate or translate while vibrating, which means our program meets the expectation. 

\begin{figure}[hbtp]
	\centering
	\includegraphics[width=.5\textwidth]{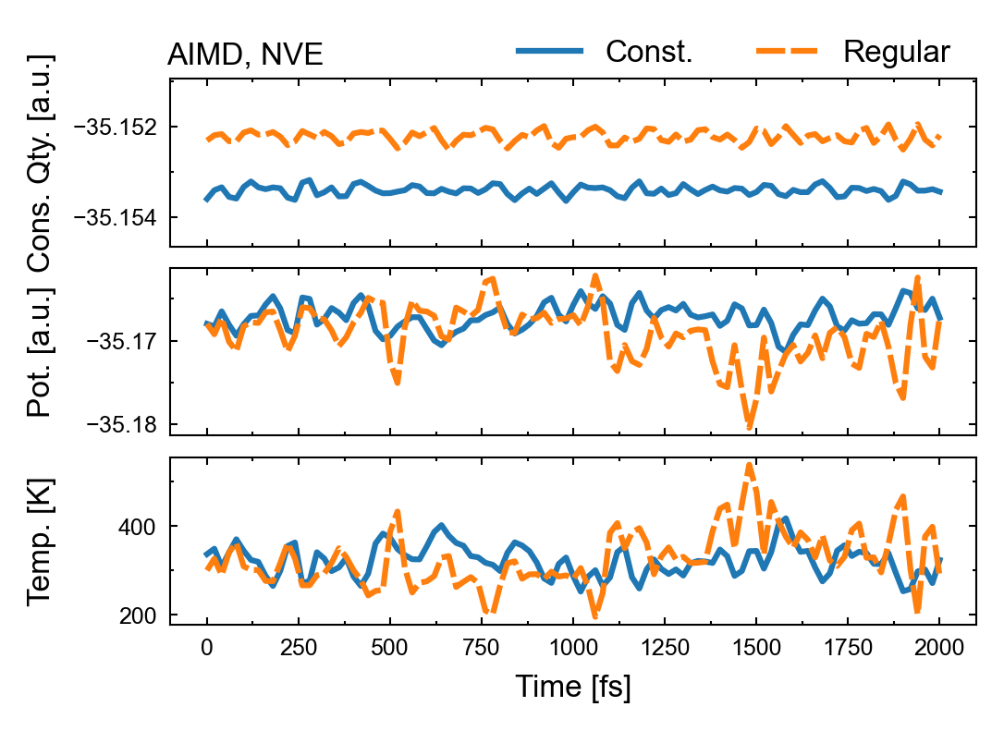} \\
	\caption{Conserved quantity (Cons.~Qty.), potential (Pot.), and temperature (Temp.) fluctuation of the AIMD trajectory of NVE ensemble with/without roto-translational constraints (Const./Regular). \label{fig:rtc_qm_nve}}
\end{figure}

\begin{figure}[hbtp]
	\centering
	\includegraphics[width=.5\textwidth]{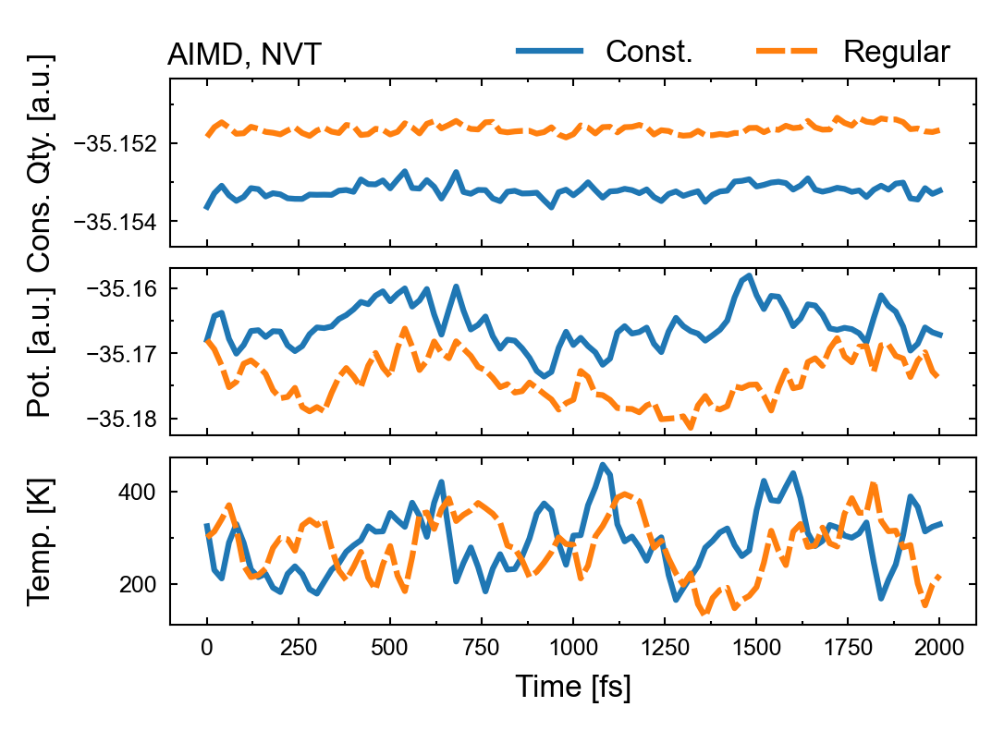} \\
	\caption{Conserved quantity (Cons.~Qty.), potential (Pot.), and temperature (Temp.) fluctuation of the AIMD trajectory of NVT ensemble with/without roto-translational constraints (Const./Regular). \label{fig:rtc_qm_nvt}}
\end{figure}

The results of statistic fluctuation is similar with rotational constraint, which is showed in Fig.~\ref{fig:rtc_qm_nve}--\ref{fig:rtc_qm_nvt}, no significant difference from the regular MD. The drift and fluctuation difference values are in Table \ref{tbl:consq}. 

There is a small difference in initial temperature between constrained MD and regular MD. This difference of the temperature is from the program implementation, not the theory, and will not affect the final equallibrium temperature. We set the initial temperature to 300~K in all trajectories, when using the roto-translational constraint, the initial temperature is always a little higher than our settings, in Fig.~\ref{fig:rtc_qm_nve}--\ref{fig:rtc_qm_nvt} is about 336 K. To explain this, we need to understand what is actually happening in the program. Typically, the constraint algorithm does nothing at the zeroth step of MD.
After initializing the atomic velocity according to the temperature, usually the atomic velocity of the system does not satisfy the constraint condition. Then we apply the constraint to the system, which will do work on the system and cause a change in the total energy. Once the constraint condition is satisfied at a certain step, the constraint algorithm will theoretically do no work on the system at subsequent steps in the MD. This will be shown as a leap in total energy only from zeroth to the first step. 

In order to avoid the total energy leap during the statistic, we used the \\ ``CONSTRAINT{\_}INIT T'' option, which forces the constraint algorithm to be applied to the zeroth step. Then the program scales the atomic velocities according to the set temperature. 
The overall COM velocity of the system is usually zero when scaling, but with the rotational-translational constraint, we fix the translational motion of two molecules, and the remaining molecules will be attracted to or moved away from each other by the interactions, resulting in a total COM velocity for the system. Then the CP2K program will also scale the total COM velocity, so that the initial temperature will always be different from the temperature we set. We have not modified this problem in order not to affect the use of other functions in CP2K. Although the initial conditions are changed, this problem does not affect the use of the thermostater. 

\subsubsection{\ch{LiBH4} crystal}
\label{sec:libh4}
 As a solid state electrolyte material, the rotation of \ch{BH4-} anions in \ch{LiBH4} crystal has been observed by experiment and simulation.\cite{libh41, libh42, libh43} In this section we take \ch{LiBH4} crystal as a test example. The system is a supercell with 216 atoms inside. 35 \ch{BH4-} anions are considered as 35 rotator to be constrained. The \ch{LiBH4} results are similar with the 3 ammonia results. The error of constraint is small. For rotational constraint simulation, the max rotational error is $\sigma_{max}= 2.34\times 10^{-6}$~\AA. For roto-translational constraint simulation, $\sigma_{max}= 2.34\times 10^{-6}$~\AA\;and $\tau_{max} = 4.41\times 10^{-6}$~\AA.

\begin{figure}[hbtp]
	\centering
	\includegraphics{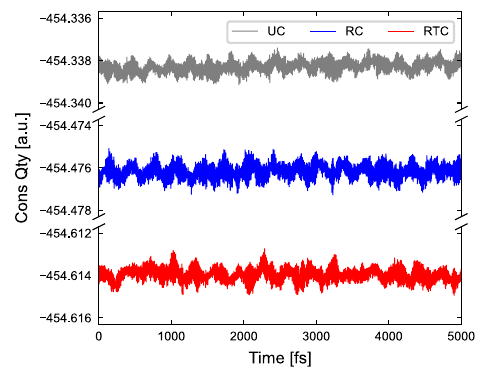} \\
	\caption{Comaprison of conserved quantity (Cons Qty.) fluctuation of unconstrained (UC), rotatinal constrained, and roto-translational constrained trajectory.
		\label{fig:libh4Comp}}
\end{figure}

All drift/fluctuation difference of conserved quantity from the no constraint MD is showed in Table~\ref{tbl:consq} and Fig.~\ref{fig:libh4Comp}. 
Like 3 ammonia system, the fluctuation difference values are little than $1\times10^{-4}$~a.u.. The absolute values of fluctuation are about $3\times10^{-4}$ which are small compare with the system energy about -450 a.u. The drift difference values showed a bit higher results in Table~\ref{tbl:consq}, but they are very small when compared with the total system energy. 

 Our conserved quantity results, both in drift and fluctuation, generally are a little higher than the Nola group's results. However, it is hard to evaluate the methods from the comparison. Many causes can affect the conserved quantity: different constraint method, thermostator, integrator, MD software, etc. We used larger system, longer statistic length and larger timestep, more DOF are constrained, while in both 3 ammonia and \ch{LiBH4} system, most drift and fluctuation values of unconstrained trajectory are in the same order of magnitude with the values of constrained trajectory, which implicates the stability of our method.
 
\section{Conclusion} 
In this work, we present a rotational/roto-translational constraint algorithm specifically designed for condensed matter simulations. The method is based on the velocity Verlet scheme, ensuring a direct constraint on velocity updates and simplifying implementation within material simulation software packages.

We successfully implemented the algorithm within a customized version of the CP2K software package. To validate its effectiveness, we conducted MD simulations using both the NVE and NVT ensembles on a system of three ammonia molecules and a \ch{LiBH4} crystal. By evaluating the drift and fluctuation of conserved quantities during these simulations, we confirmed that the algorithm can selectively constrain the rotational/roto-translational motions of molecules, enabling stable long-term MD simulations.

This algorithm's capabilities make it a versatile tool for studying various rotation-related constrained sampling and phenomena in condensed matter, including the paddle-wheel mechanism in solid-state electrolytes. 
In the future, we plan to apply this method to investigate solid electrolyte systems and enhance the code structure and user interface for broader accessibility.
\section*{Supplementary Material}
	The supplementary material includes theoretical derivation of roto-translational constraint, fluctuation of quantities with FFMD, and animates of MD trajectory results. The source code can be found in https://github.com/jtyang-chem/rconst\_cp2k.

\begin{acknowledgments}
This research was sponsored by the National Natural Science Foundation of China (grants 22073035, 22372068, and 21773081), and Graduate Innovation Fund of Jilin University. We also thank the Program for the JLU Science and Technology Innovative Research Team.
\end{acknowledgments}
\section*{Data Availability Statement}
The data that support the findings of this study are available
from the corresponding author upon reasonable request.
\bibliography{manuscript}

\end{document}